\documentclass{scrartcl}

\usepackage{amsmath,amssymb,graphicx,booktabs,subfig,caption,float}
\usepackage[colorlinks=true,citecolor=blue,linkcolor=black,urlcolor=blue]{hyperref}
\usepackage[backend=biber,style=authoryear,sorting=none]{biblatex}
\usepackage[nameinlink]{cleveref}

\title{A Theory of Value Growth}

\author{Wang, Zhuo\thanks{Research Institute of Economics and Management, Southwestern University of Finance and Economics, Chengdu, China. Email: \texttt{zwang@swufe.edu.cn}. This paper is my original work. I thank my colleagues and friends for their comments and suggestions, also thank ChatGPT for excellent writing assistance. This research received no funding.}}

\date{September 28, 2026}

\begin{document}
\maketitle

\begin{abstract}
Classical growth theory places physical productivity at the center of economic growth. In modern service-based economies, however, growth increasingly means creating higher-valued products rather than increasing output quantities. This paper develops a theory of value growth based on a socially and institutionally sustained value scale. The framework unifies value creation, finance, and distribution and offers an account of major patterns in US economic history. The proposed framework is intended as a new approach to macroeconomic analysis and policy making.

\end{abstract}


\section{Introduction}
An economic theory offers a highly abstract account of the economic process that explains its most important features. It also provides guidance for economic policy toward its intended outcomes. However, the dominant Neoclassical and Keynesian theories fail these tasks in the current economic situation.

Mainstream economics teaches us that higher living standards depend on greater productivity, that is, more output per worker. This framework works well in an economy built on agriculture and manufacturing products. However, it falls short in modern economies, which primarily rely on services. The quantity of services a barber, doctor, or consultant can provide has remained stable over time. Progress in these sectors often means more valuable services, rather than more quantities per worker. The concept of productivity becomes even more awkward in the world of the digital economy. A piece of music, a movie, or an online subscription can generate enormous value by reaching millions or billions of users. Yet its producers are hardly more ``productive'' in the traditional sense than those with smaller audiences.

Conventional theory attributes underdevelopment to low productivity. That is only part of the story. Development means not only producing more but also gaining access to higher valuation regimes through brands, platforms, culture, standards, and even geopolitical control. Nominal income divergence partly arises from the value hierarchy itself, rather than physical productivity alone. 

Another fundamental problem in classical growth theory is the peripheral role of money and finance. The central mechanism in neoclassical growth theory---saving and capital accumulation---is obsolete. Modern economies channel resources through financial \textit{tokens} that financial institutions create \textit{ex nihilo}. They no longer rely on real saving or hard currency. Monetary and fiscal channels can supply additional tokens to support unbounded value growth, even when physical constraints limit output quantities.

This paper develops a theory of value growth, and a broader account of the economy, around an externally given value scale. Prices are reflections of these values and are also subject to inflation and supply-and-demand conditions.
For reasons I will explain later, I don't think the predominant approach---treating prices as the equilibrium outcomes optimally derived from demand and supply curves---addresses the kernel of price phenomena. Taking the underlying valuations as given makes it straightforward to examine how production and finance turn them into incomes without much derivation. Growth means climbing the value scale and creating higher-valued output. Positions that command larger value-chain markups can support higher wages and profits.

Growth in the value scale requires proportional growth of money and credit provision, given price stability. This means monetary elements are as important as real production. In practice, financial conditions are shaped by various monetary and fiscal policy regimes. The relationship between the policy regimes and the real production process---and a possible mismatch between them---produces distinct growth patterns that reveal themselves in history.

I will show that the relationships implied by the theory work well to explain much of US economic history. For example, if the central bank is targeting price stability and full employment, which delivers moderate inflation and declining interest rates, the national debt ratio must rise as debt grows faster than output, as was the case before the Global Financial Crisis. If rising interest rates are accompanied by a higher debt ratio, usually under a strong fiscal impetus, inflation is an inescapable outcome, as in the 1970s and the post-pandemic era. 

The theory also integrates distribution as a natural outcome of production and growth. The experience of developed economies over the past few decades---rising wage inequality and a relatively stable labor share (excluding depreciation)---is simply the consequence of an exponentially growing value chain.

Since the Lucas critique, the prevailing approach has modeled economic mechanisms as rational agents' constrained optimization problems and observed outcomes as their optimal solutions. I suspect this approach adds more complexity than clarity to economic analysis. Agents choose freely and may or may not act rationally; policymakers also wield considerable power over economic outcomes. Observed economic outcomes may be nowhere near optimal. There is little reason to believe unaffordable education and healthcare costs maximize social welfare simply because the market produces them. The narrative that globalization driven by capitalists' interests delivers the best economic outcomes is also obsolete. Even the most favorable outcomes promised by free markets depend on appropriate institutional settings and legal frameworks. Economic theories that promise rational and optimal outcomes merely by matching supply and demand curves should belong to the past. Instead, a good theory should identify economic laws that hold under the most general conditions. Such a theory should help us better understand how the economy works and inform policymakers about the trade-offs involved in their policy choices.

The rest of the paper is organized as follows. The first two sections introduce the value scale and a multi-stage structure of production as the starting point of the theory. The next section introduces finance to the theory and derives the finance-value identity. The subsequent analysis verifies the theory empirically through data and history. The final section discusses wage distribution and labor’s share in light of the value chain.

\section{The Value Scale}

Assume $N$ types of goods or services with the following given value scale:
\[
v_{1}<v_{2}<\cdots<v_{N}.
\]
Low-value goods may include necessities and standardized manufactures. High-value goods may include intellectual property, brands, advanced services, and cutting-edge technologies.

Observed prices reflect underlying values, the general price level, and supply and demand conditions:
\[
p_{i}=pv_{i}x_{i},
\]
where $p$ denotes the general price level, $x_i$ captures ordinary market-clearing variation, and $v_i$ captures a persistent valuation position.

\paragraph{Why a value scale?}

Markets are always embedded in institutions, cultural norms, and social meanings. Given the social context, human beings naturally understand that some goods are more valuable than others. A value scale is a simple way to capture this understanding. It is not a physical measure, but a social and institutional one. 

Perhaps the greatest obstacle to economic theorizing is the habit of imagining an economy determined by sets of supply and demand curves. The fact is, prices depend on supply and demand, but supply and demand also depend on prices. In many settings, stable curves simply do not exist.

Yet supply and demand curves are analytical tools; they need not exist as concrete objects. But treating observed prices as somehow optimally determined by these curves obstructs an effective explanation of important price phenomena.

Consider an ordinary T-shirt and a luxury branded one. They may have similar production costs but very different price tags. Aesthetics, branding and cultural relevance give them different values. Economics professors may argue these factors are behind the curves. But everyone would agree that thinking about demand and supply curves does little to help us understand the price difference.

The \emph{Journal of Financial Economics} charges an extraordinary \$850 for your submission to the journal. It is not about production cost. Nor do supply and demand give a satisfactory explanation. The journal's quality and the dominant position of American academia give it authority and make it highly valued among scholars. This is value generated by authority.

A Boeing aircraft might exchange for millions of ordinary T-shirts. Production costs certainly differ. But production costs, including labor wages and intermediate goods, also depend on the price the final product can command. An explanation with only supply and demand risks a tautology. Aircraft occupy a higher value position because of their technological complexity and the producers' monopolistic position.

The secret of economic advancement is the ability to produce higher-valued goods and services---through brands, culture, authority, technology and other means. The higher the value of the final products, the higher the income they return to the means of production. This is the source of higher living standards. This is the reason why American workers enjoy greater purchasing power than Kenyan workers.

The claim that trade makes both parties better off is popular. But trade alone does not produce economic advancement. Trading goods at the bottom of the value hierarchy does not deliver a higher living standard. Global value chain studies point in the same direction: higher-income countries generally occupy higher-value positions. Low-income countries catch up by moving their exports up the value chain.

\paragraph{What is value?}

Value has a highly abstract meaning here. It is easy to imagine that the Mona Lisa is more valuable than a kilogram of oranges, but it is hard to explain analytically. Value serves as a conceptual device rather than a directly measurable quantity. It is not identical to an item's price tag, although prices are observable and separating underlying value from price is difficult in practice. 

A ChatGPT subscription priced at \$20 per month may cost less than a haircut. This does not establish that ChatGPT is less valuable. 
Higher-value goods usually build on lower-value ones. ChatGPT, for example, is built with programming languages and trained on GPUs. It therefore occupies a higher position in the value chain than these inputs. 

But technology is not the only source of the value advantage. A movie can be higher-valued than ChatGPT if it is highly popular among audiences. Financial services would be more valuable in a bull market. If we are keen on a quantity, loosely speaking, the (social) value of a product is the aggregation of all consumers' subjective valuations. 

The approach departs from classical economics by treating growth primarily as value creation rather than as improvements in physical productivity. Value need not correspond to a physical quantity. Physical goods can be valuable; so can a piece of advice or an online subscription. This is especially true in a service-based economy. ChatGPT exemplifies new value creation at the technological frontier. Its value does not stem from traditional productivity---OpenAI employees do not necessarily produce more lines of code per hour than other programmers. Rather, it stems from the new technology and the intelligence embedded in the product, which make it highly valued by a broad audience. This process of value creation, rather than productivity in the traditional sense, is the central driver of modern economic growth.

\paragraph{Not another theory of value.}

Taking $\{v_i\}$ as exogenous does not mean that values are literally external to the economy or constant. We may write: $v_i=\mathcal{V}(\mathbf{z}_i)$, where $\mathbf{z}_i$ may include brands, intellectual property, market power, authority, consumer tastes, and all other influences. The debate over what determines value has a long history, from objective to subjective value theories, and from labor value to marginal utility theories. This paper does not enter that debate. 

The value scale is admittedly a substantial conceptual simplification. In practice, values may be multidimensional and context-dependent: a good may be highly valued in one context but less so in another. The model abstracts from these nuances because they do not affect the substance of the theory.

The purpose of this paper is to develop a theory of growth and finance given the existence of such a value scale. The value scale is an analytical starting point, just as a utility function is for many existing theories. 

\section{The Structure of Production}

Production is conceptualized as a multi-stage process along the value chain: a sequence that transforms lower-valued inputs into higher-valued outputs. At each stage $i$, a monopolistic firm produces output valued at the $i$-th position in the value scale. Abstracting from supply and demand fluctuations, the output has a price of $pv_i$, where $p$ is the general price level.

The chain starts at $i=1$, with $v_1>0$. Stage 1 uses original factors of production rather than an intermediate good from a preceding stage. For $i\geq2$, firm $i$ combines $q_{i-1}$ units of the preceding good with $n_i$ units of labor, paid at wage $w_i$, to produce $q_i$ units valued at $v_i$. Its value added is
\begin{equation}
\text{VA}_{i}=pv_{i}q_{i}-pv_{i-1}q_{i-1}.\label{eq:value-added}
\end{equation}
Profit equals value added less wages:
\begin{equation}
\pi_{i}=pv_{i}q_{i}-pv_{i-1}q_{i-1}-w_{i}n_{i}.\label{eq:profit-1}
\end{equation}
For the first stage, $\text{VA}_1=pv_1q_1$ and $\pi_1=pv_1q_1-w_1n_1$. The absence of fixed capital may seem strange. But this theoretical issue can be easily overcome by considering the services provided by fixed capital as intermediate inputs.

This structure has natural parallels in reality. Consider the production of computers: pure silicon is made from sand, which is further sliced into wafers; then transistors are fabricated, packaged into CPUs, and mounted on motherboards to make computers. Real-world production can be much more complex than a simple chain. In fact, the more sophisticated the economy, the more complex the production network. A computer may use thousands of components, each with its own production chain. 

The essential point is: production is not a homogeneous process. Capital deployed in agricultural production and chip manufacturing is not comparable. Economic advancement does not emerge from homogeneous capital accumulation---accumulating more agricultural capital does not automatically lead to chip-manufacturing capabilities. Nor does the diminishing return of capital make more sense in this context. The stages of production along the value chain capture this heterogeneity. The ability to produce at the higher end of the value chain usually comes from technological advantages, patents, brands and so on.
 
The stages of production might remind us of the Austrian school's ``roundabout'' production, but the interpretation here differs. More stages need not mean a longer production period as the Austrian school assumed; they may operate simultaneously. Higher stages use lower-stage outputs as inputs, but they need not be closer to final consumption. Consumers can consume goods from any stage. Ultimately, the chain represents a hierarchy of value creation, not a technical sequence of production.

\subsection{Wage Inequality}

To simplify the notation, assume one-to-one transformation: with proper normalization of units, one unit of intermediate input corresponds to exactly one unit of output, so $q_i=q_{i-1}=q$. Also assume a linear labor technology, $q_i=a_i n_i$, where $a_i$ is labor productivity. Profit becomes
\begin{align}
\pi_i&=pq(v_i-v_{i-1})-w_i n_i\nonumber\\
&=q\left(p\Delta v_i-\frac{w_i}{a_i}\right)\nonumber\\
&=pq(\Delta v_i-\omega_i),\label{eq:profit-2}
\end{align}
where $\Delta v_i=v_i-v_{i-1}$ for $i\geq2$, and $\Delta v_1=v_1$. $\Delta v_i$ denotes the value-chain markup; $\omega_i=w_i/(pa_i)$ is real unit labor cost. Viability requires $\pi_i\geq0$, or
\(\Delta v_i\geq\omega_i.\)
That is, the value-chain markup must cover real unit labor cost.

If labor receives a share $\lambda_i$ of value added, then
\[
w_i n_i=\lambda_i\text{VA}_i=\lambda_i pq\Delta v_i.
\]
The real wage is therefore
\begin{equation}
\frac{w_i}{p}=\lambda_i\frac{q}{n_i}\Delta v_i=\lambda_i a_i\Delta v_i.\label{eq:real-wage}
\end{equation}
Workers earn more when productivity is higher, their firms command larger value-chain markups, or labor receives a larger share relative to profit earners.

Holding productivity and labor's share constant, industries with greater value added enjoy higher wages and profits. Different theories would suggest different distributions of $\Delta v_i$. One possibility is that $\Delta v_i$ rises with $i$. That is, higher-end production not only produces more valuable output but also produces higher value added. For example, finance and consulting produce higher value added than manufacturing. If this is the case, wage inequality and concentration of profits are natural consequences as the value chain extends. The shape of the value chain is therefore at the center of the distribution problem. We will come back to this point in the last section.

\subsection{The Inflationary Process}

Conventional wisdom favors moderate inflation, often around 2\%: prices should rise steadily enough to encourage consumption and investment, but not so fast as to cause disruptive changes. Consider a steadily rising $p$. Assume nominal wages keep pace with inflation, leaving real wages unchanged. Recall that
\[
\pi_i=pq(\Delta v_i-\omega_i).
\]
For given quantities and real unit labor costs, firms with larger value-chain markups record larger absolute increases in nominal profit.

The same applies to nominal wages. Higher-value-added industries deliver larger absolute wage gains during inflation. Higher earners start with more and gain more in monetary terms. Uniform price inflation therefore widens nominal income gaps. This statement concerns absolute nominal differences under the assumed constant real wages.

So far, input and output prices refer to the same period. In practice, firms may buy inputs before selling their output, and prices can change in between. Suppose production takes one period, longer than price adjustment. Firms purchase inputs and labor at price level $p$ and sell output at $p'$. Then
\[
\pi_i=pq\left(\frac{p'}p v_i-v_{i-1}-\omega_i\right).
\]
If $p'>p$, inflation raises nominal profit. The gain is proportional to $v_i$. So higher-end firms enjoy more inflation gain. The rest of the paper assumes the production period is short enough so that input and output price changes are contemporaneous.

\subsection{Sources of Nominal Growth}

We may, but need not, assume that only the last stage produces final goods or services. If all wage and profit income is used to purchase final goods and services, the economy has a closed circular flow. Total income satisfies:
\begin{align*}
Y&=\sum_{i=1}^N(w_i n_i+\pi_i)\\
&=pq\sum_{i=1}^N(v_i-v_{i-1})\\
&=pq(v_N-v_0).
\end{align*}
Assume $v_0=0$, since stage 1 purchases no preceding intermediate good. Total income thus equals nominal final output:
\begin{equation}
Y=pqv_N.\label{eq:output-decomp}
\end{equation}

In this case, $Y$ would correspond to nominal GDP. Taking logarithmic growth rates over time, denoted by hats, gives
\begin{equation}
\hat Y=\hat p+\hat q+\hat v_N.\label{eq:growth-decomp}
\end{equation}
Nominal GDP growth thus has three sources: inflation, quantity growth, and value-chain expansion.

\paragraph{Taking it to the data.}
Consider the US economy as an example. 
Nominal GDP growth and inflation are readily available. However, a direct measure of $q$, the quantity of goods and services, is not. The decomposition requires a quantity measure rather than ``real output'' obtained by deflating nominal output. For this purpose, I construct a quantity index:
\[
Q_{\text{index}}=w_t^I I_t+w_t^S H_t+w_t^C C_t.
\]
Here $I_t$ is the Federal Reserve's industrial production index, a proxy for physical output. Its weight, $w_t^I$, sums the GDP value-added shares of manufacturing, mining, and utilities published by the Bureau of Economic Analysis (BEA).

There is no service quantity index $H_t$ compiled from the number of patients treated in hospitals, the number of students educated in schools and so on. I use BLS' index of aggregate weekly hours of private service-providing employees to approximate service quantities. This assumes roughly constant service quantities per labor hour in the service sector: a barber can give only so many haircuts and a doctor can treat only so many patients. Violations of this assumption are discussed later.

Construction appears in neither industrial production nor private service hours, yet it is an important component of output. The index of aggregate weekly hours of all construction employees proxies $C_t$, with construction's GDP value-added share as $w_t^C$. The remaining weight is $w_t^S=1-w_t^I-w_t^C$. Growth in the composite index measures aggregate quantity growth; subtracting population growth gives per capita quantity growth.

The final value $v_N$ is unobservable. The residual in \Cref{eq:growth-decomp}, after subtracting inflation and quantity growth, serves here as an indicator of value growth. Admittedly, it is only an approximation, subject to all sorts of measurement errors.

Critics may question the use of labor hours to approximate service output. Certainly, labor productivity does not change for some service sectors. But there are exceptions. Consider a piece of music, a movie, or a mobile app. These products may differ enormously in their values, depending on their popularity, despite being produced by teams of similar sizes. But this cannot be interpreted as higher productivity in providing the service. This exactly exemplifies value creation. The popularity depends on creative quality, consumer preferences, cultural relevance, and even luck. A producer does not decide: should we manufacture another million view-hours? Instead, they decide: what should we create that people will value enough to watch, use, recommend or pay for? Therefore, the residual is better considered as a proxy for value creation rather than a proxy for productivity. 

\begin{figure}[tbp]
\centering
\includegraphics[width=\textwidth]{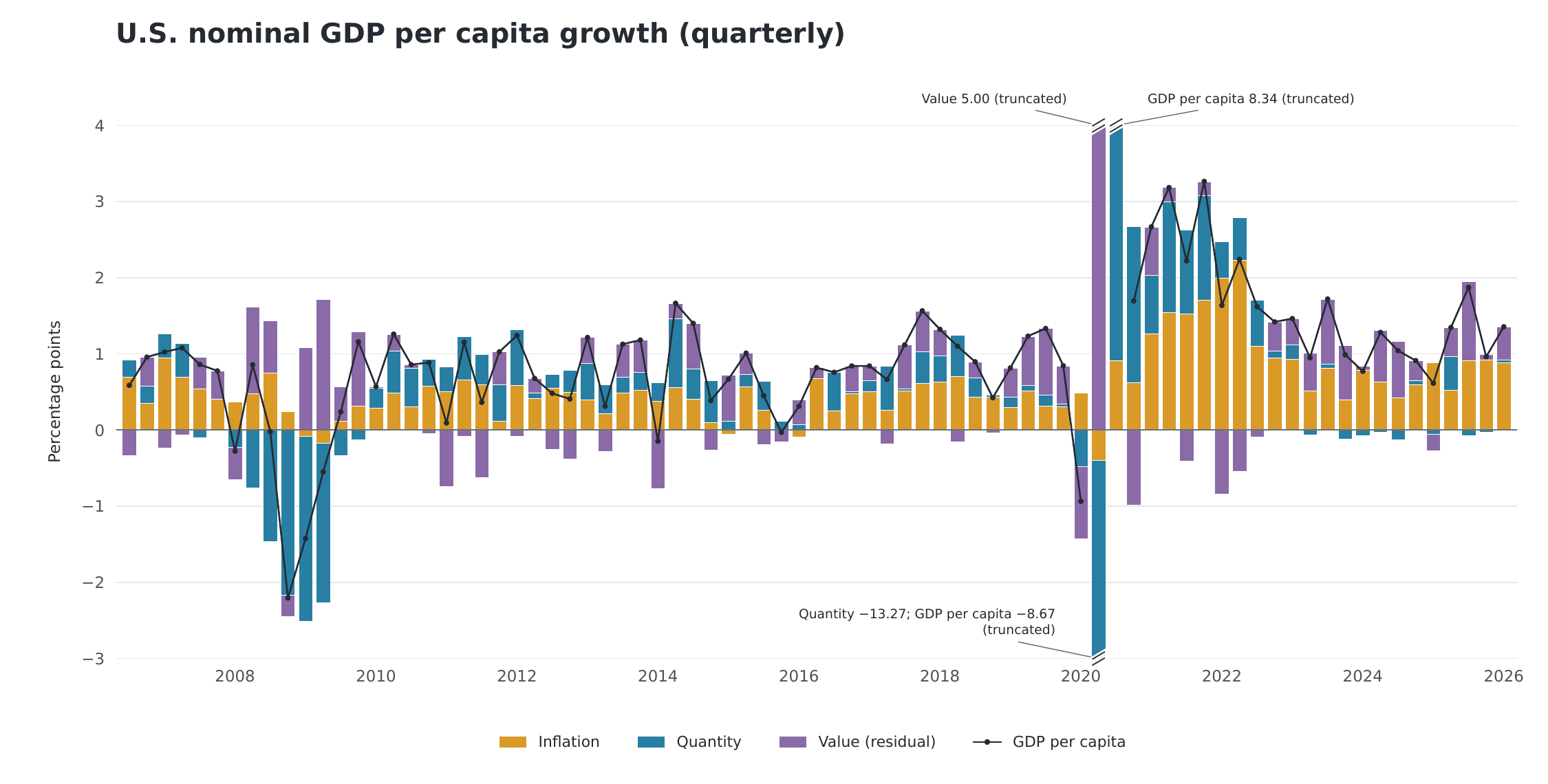}
\caption{Decomposition of U.S. Nominal GDP Growth per capita}
\label{fig:growth-decomp}
\begin{minipage}{\textwidth}\small
\textit{Notes:} Growth is calculated as annual log changes. Contributions are in percentage points. Data are sourced from FRED. Inflation is computed from the GDP deflator. The quantity index is computed from the Fed's industrial production index, BLS' index of aggregate weekly hours of private service-providing employees, and the index of aggregate weekly hours of all construction employees. The weights are their value-added shares in GDP accounting. 
\end{minipage}
\end{figure}

\Cref{fig:growth-decomp} presents the decomposition from 2006:Q3 to 2026:Q1. Quantities fall sharply during the GFC and COVID crisis, offset by positive value growth (the residual). For example, during the COVID crisis, fiscal injections helped fuel the boom in the stock market, which created positive values but fewer physical quantities. Excluding crisis episodes, price growth accounts on average for 52\% of nominal GDP per capita growth, quantity growth for 34\%, and the residual interpreted as value growth for 14\%.

Thus, real quantity growth plus inflation do not provide a complete account of the observed nominal growth. The decomposition, however, cannot rule out unmeasured productivity or measurement error. The argument rests on the conceptual distinction between quantity and value; the numerical decomposition is only illustrative.

\subsection{Summary}

Growth is not simply the production of more things. It is also the ability to create goods and services that society values more highly and to claim a better position in the value hierarchy. This distinction matters in economies built on services, digital products, brands, intellectual property, and institutional authority. A value scale provides a starting point for analyzing these economies without utility maximization and supply and demand curves.

\section{The Role of Finance}

The salient feature of modern finance is that real production resources are allocated by fictitious financial tokens generated by financial institutions. Any theory that theorizes modern capitalism cannot miss this feature. As Schumpeter once said: “...it has to be recognized that essential features of the capitalist process may depend upon the ‘veil’ and that the ‘face behind it’ is incomplete without it” \parencite[265]{schumpeter1954}.

The term ``financial token'' is borrowed from the Crypto world. I prefer the word ``token'' to signify that it is issued freely and carries the notion of value. It can be fiat money, credit, or any security that grants purchasing power. This paper abstracts from different channels of finance, as it is not a monetary paper that tries to explain how modern finance works. However, three features of modern finance are relevant here.

First, investment never needs ``real savings'', or any pre-saved resources. For a firm borrowing from a bank, the bank evaluates the return and risk and issues tokens out of thin air, which grant the firm purchasing power. 

Second, financial tokens respond elastically to demand. There is no limit to the total amount of tokens that can be created unless the authority imposes one. Contrary to the common belief that only the central bank prints money while other financial institutions merely borrow existing funds, any financial institution creates money, where the central bank controls the difficulty of their money creation.

Third, although tokens are issued \textit{ex nihilo}, their issuers charge interest. The interest rate is treated as a policy variable. This is the case where policy not only controls the short-term but also the long-term interest rates. It is not yet the reality, but the world is clearly moving toward it. There is no loanable funds market that determines the real interest rate. There is no natural rate of interest that acts as the gravitational force for the interest rate to be pinned on. Interest rates are solely controlled by the monetary authority subject to its tolerance of inflation and debt levels. 

Let $M$ denote the total financing required by the production process and $r$ the uniform interest rate. Actual rates vary with risk premia, which the model abstracts from. Total interest income is therefore $rM$. It does not matter who owns the financial resources---be it financial institutions, wealthy capitalists, or the state. Nor does it bother us whether the prevailing interest rate maximizes anyone's utility. The critical assumptions are: financing responds elastically to production demand, subject to policy constraints, and interest income is distinct from wages. 

\subsection{Production with Borrowing}

From this point onward, let $q=n=1$, to abstract from the quantity of output and focus only on the value aspect. Assume firms have no retained earnings and borrow the full cost of intermediate inputs and wages. Firm $i$ thus borrows
\[
I_i=pv_{i-1}+w_i,
\]
defined as its investment. It repays principal and interest when production ends, earning profit
\[
\pi_i=pv_i-(1+r)(pv_{i-1}+w_i).
\]
Assuming zero entrepreneurial profit gives
\begin{equation}
\frac{pv_i}{1+r}=pv_{i-1}+w_i.\label{eq:zero-profit}
\end{equation}

Zero entrepreneurial profit is a distributional assumption, not an appeal to perfect competition. \footnote{Perfect competition never exists. Every product differentiates itself from others; even a small grocery store has a local monopoly over its particular location and convenience. Capitalism lives and dies with profit. A perfect competition assumption that erases all profits should not be the baseline assumption for capitalism.} It is equivalent to assuming financial providers capture every surplus entrepreneurs produce, leaving only normal managerial income, counted as wages.

For $i\geq2$, nonnegative wages require
\[
\frac{pv_i}{1+r}\geq pv_{i-1},\quad\text{or}\quad
r\leq\frac{\Delta v_i}{v_{i-1}}.
\]
A firm's interest rate cannot exceed its proportional value increment. 
For $i=1$, $w_1=pv_1/(1+r)>0$. That is, the producers of the original means of production are always rewarded.

For the entire chain to remain viable, the rate must respect the ceiling
\begin{equation}
\bar r=\min_{i=2,\ldots,N}\left\{\frac{\Delta v_i}{v_{i-1}}\right\}.\label{eq:rate-ceiling}
\end{equation}
For $N\geq2$, viability requires $r\leq\bar r$; equality gives zero wages at a binding stage. Therefore, higher rates place the greatest pressure on firms with small proportional value increments. A rate above the ceiling breaks the chain, eliminating its weakest links first.

\subsection{The Aggregate Identity}

Total financing across firms is
\begin{align*}
M&=\sum_{i=1}^N I_i
=\sum_{i=1}^N\frac{pv_i}{1+r}
=\frac{p}{1+r}\sum_{i=1}^N v_i.
\end{align*}
Define $V=\sum_{i=1}^N v_i$ as gross value across all production stages. Then
\begin{equation}
(1+r)M=pV.\label{eq:rmpv-identity}
\end{equation}
This is the \emph{finance--value identity}. The right-hand side is gross nominal value from production; the left-hand side is the gross value of financial tokens inclusive of interest. The identity connects financial activity with the real production activity. Note that $V$ is gross output value of all production stages, and $pV$ is its nominal value; neither is restricted to final output such as GDP.

The theory implies: holding value growth fixed, rate cuts require a sufficient increase in credit growth to raise inflation; rate increases require a sufficient reduction in credit growth to lower it. If the change in credit offsets the change in the interest rate, the price level will not change.

Nonconventional though it is, the intuition is clear: a higher interest rate means a higher cost of production. Under the monopoly assumption, firms would raise the price to meet the higher cost. The causality could also run the other way: a higher price level means a higher cost of production, which requires more financing to cover the cost, which in turn puts upward pressure on interest rates. The identity does not determine the direction of causality, but it tells us the laws that govern the relationship between the variables.

This interpretation implies an unusual policy combination. Lower rates paired with a shrinking central-bank balance sheet to restrain credit growth may appear contradictory from a Keynesian perspective, yet their combined effect can reduce inflation by slowing financing growth relative to value growth. This is especially relevant to the post-pandemic United States. For a country facing deflation, such as China, monetary easing may fail to reflate the economy unless strict controls on local-government debt also ease.

\Cref{eq:rmpv-identity} is not like anything we have seen in economics, yet it is the fundamental equation that underpins the theory. So it deserves more explanation.

\paragraph{Links to the quantity theory of money.}

If $r$ is held constant, the equation resembles the quantity theory of money (QTM).\footnote{The quantity theory of money generally states that the price level of an economy is determined by the money in circulation. The iconic equation is $MV=PY$, where $M$ is the money stock; $V$ is the velocity of money; $PY$ is the nominal output or nominal GDP.} Although QTM is one of the most famous identities in economics, the theory is never well defined. 

For example, which monetary measure belongs in the equation: currency, high-powered money, or deposits? QTM seems to link the transactions needed by the economy with the money supply. But transactions can take place without exchanging against money. Business transactions are often done with commercial paper. In my theory, $M$ has a different meaning. It is the total financing needed to support production, which is not necessarily the same as the money supply.

The QTM uses final output GDP in the equation, which is also problematic. Money finances intermediate transactions as well as final purchases. The financing needed to support a given final output depends on production structure. Separate factories exchange intermediates repeatedly; a vertically integrated giant factory can dispense with many of these transactions and require less money. So transaction needs should consider the entire production structure, not just final output. 

The velocity of money is also difficult to interpret. The frequency with which money ``changes hands'' is an awkward description of a system in which money largely consists of entries in bank computers. In practice, the velocity of money often becomes whatever residual makes the equation hold. In \Cref{eq:rmpv-identity}, production structure and the interest rate help determine the relationship that velocity summarizes.

\paragraph{Links to Gibson's law.}

The identity permits a positive relationship between the interest rate and the price level. This is Gibson's law: interest rates correlate positively with the general price level, rather than its rate of change. Gibson explained it as: Lower prices reduced living expenses, leaving more income available to save. Greater saving increased the supply of loanable funds, pushing interest rates down.\footnote{Gibson's law is unfortunately named “Gibson's paradox” by J.M. Keynes, because it cannot be explained by the Keynesian theory. But it is not a paradox, and should not be called a paradox. The correlation of the interest rate and the price level discovered by Gibson is robust across many historical episodes and is still relevant today.} 

Despite its recurrence across historical periods, prevailing theory has struggled to explain it.\footnote{\textcite{barsky1988} explain Gibson's law under the gold standard. However, evidence shows this correlation extends beyond the gold standard. } The dominant theory regarding the interest rate and price changes is inspired by \textcite{wicksell1936} and \textcite{keynes1930}, which emphasizes a negative relationship between interest rates and inflation.

\Cref{eq:rmpv-identity} confirms Gibson's law, and it applies beyond the gold standard. Whether markets or policies determine $r$ and $M$, the identity constrains their relationship with output value and prices. Those constraints define policy trade-offs, which will be examined in later sections.

\paragraph{Is Keynesian theory wrong?}

Since Keynes published his \textit{General Theory} \parencite{keynes1936}, it has become the dominant framework for understanding short-run macroeconomic fluctuations, and a basis for policy prescriptions to stabilize output and inflation. The theory advanced here implies very different policy prescriptions from the Keynesian theory, which deserves comment.

First, Keynes assumes it is ``real'' variables that matter in economic life, whereas the thesis of this paper is that we fundamentally live in the nominal world. Nominal value creation matters as much as real productivity gains. If economic advancement is mainly about value creation, the real income or return is hard to define. 

Second, the essential Keynesian argument is: full employment requires efficient aggregate demand, which is largely driven by investment, while the latter is determined by the difference between ``marginal efficiency of capital'' and the interest rate. When the former is greater than the latter, there is an incentive for entrepreneurs to expand. Otherwise, the economy would suffer from insufficient investment and involuntary unemployment.

In my view, this argument seems to build on a foundation that conflates routine production of capital goods (also called investment) with investment in new business opportunities. Economies continuously produce intermediate goods to sustain production, which differs little from consumption-goods production. Given the institutions and production structure, established practices largely determine production and employment, in both capital-good and consumption-good production sectors, with only marginal adjustments to changing circumstances. Unless rates become destructively high, firms generally adapt prices or quantities rather than quit. Without a major external shock or a severe mismatch in production structure, there need be no shortage of investment or effective demand whatever the saving rates are. That's why countries with very different saving rates can sustain full employment without problems.

Keynes's account better describes a special case: a boom in capital accumulation driven by new opportunities, followed by a realization that high expected returns cannot materialize without sufficiently low interest rates or additional demand, perhaps from the state. The railroad investment boom, the dot-com bubble, and the current AI hype illustrate this type of investment wave. 

Third, Keynes largely treats capital as homogeneous and employment as a quantity. Accumulating capital then brings diminishing marginal returns, which would require ever lower interest rates to maintain investment levels. As discussed earlier, capital is heterogeneous. So is labor. It makes little sense to talk about employment as a quantity without discussing what types of occupations and skills the economy needs. Capital accumulation also does not necessarily mean diminishing returns insofar as value is concerned. For there is always an opportunity to create goods or services that command higher nominal values. Nor is there a long-run tendency for the interest rate to trend towards zero. \footnote{Thinking about the situation we are in today justifies this: the economic buoyancy created by the large fiscal injections during COVID completely reversed the long-run trend of ``secular stagnation''. Businesses are more profitable, interest rates are higher, investment is booming, and the stock market is at an all-time high, though inflation offsets part of the nominal gains. The monetary- and fiscal-policy-engineered boom is real, which has little to do with real capital and productivity.}

Fourth, the liquidity preference function Keynes proposed completely rules out the case where higher (lower) interest rates could coexist with higher (lower) money supply. The authority controls either money supply or the interest rate; the other one follows from the liquidity preference function. I doubt the existence of such a function. Money growth can accelerate while rates remain high because of fiscal injections or international capital flows. Authorities can also lower rates while macroprudentially stabilizing money supply to safeguard against a debt crisis.

\subsection{Empirical Evidence}

The theory advanced here departs sharply from conventional wisdom and demands an empirical examination. This section presents a preliminary test of \Cref{eq:rmpv-identity} using the US data.
However, readers should be aware that the model's variables are conceptual, and its simplifying assumptions preclude an exact mapping into observed aggregates. The empirical question is therefore whether the evidence agrees with the directional implications, rather than matches the identity exactly.

The financing resources represented by $M$ are not directly measurable. New bond issuance or new loans alone would miss production financed through debt rollover and existing obligations. Total outstanding debt is not an exact measure either: its coverage and maturity structure differ from the financing required over a particular production period. The debt stock nevertheless provides a practical approximation because outstanding debt continues to finance assets or activities. 

Specifically, the proxy for $M$ is total debt of all sectors from the \emph{Financial Accounts of the United States}, including financial and nonfinancial debt. This raises a potential double-counting concern because financial-sector credit funds real-sector borrowing. The theory includes financial services in the value chain, so the proxy retains financial-sector debt.

The 10-year Treasury yield proxies $r$ because production investment often relies on long-term financing. Treasury yields provide a benchmark rather than firms' actual borrowing rates, which include risk premia. A well-designed empirical test should take this into consideration. But the analysis here abstracts from these nuances to keep it as simple as possible.

The value concepts are highly conceptual. The closest available proxy for $pV$ is BEA's nominal gross output of all industries, including final and intermediate output. The data do not allow $p$ and $V$ to be separately identified because the BEA's definition of real output differs from the concept of value. Quarterly gross output is available from 2005 and annual data from 1947, so the analysis uses separate annual and quarterly regressions.

To avoid spurious regression with nonstationary time series, and also to mitigate concerns about discrepancies between the available data and the theoretical concepts, the regression uses the growth-rate form of \Cref{eq:rmpv-identity}:
\[
\Delta\ln Y_t=\alpha+\beta_1\Delta r_t+\beta_2\Delta\ln M_t+\epsilon_t,
\]
where $Y=pV$ denotes nominal gross output. The specification also uses the approximation $\Delta\ln(1+r_t)\approx\Delta r_t$.

\Cref{tab:regress} reports the results. Severe recessions, including the GFC and COVID crisis, may disrupt usual relationships, so columns (2) and (4) include recession dummies. Both coefficients have the signs predicted by the theory in the annual and quarterly samples. The explanatory power is also substantial. For 1948--2025, the annual specification explains about 48\% of the variation, rising to 65\% with recession dummies. The quarterly specification has a lower $R^2$ in a sample containing two severe crises. With recession dummies, it explains about 52\% of the variation. Given the model's simplicity, these results appear supportive.

\begin{table}[!htbp]
\centering
\begin{tabular}{@{\extracolsep{5pt}}lcccc} 
\\[-1.8ex]\hline 
\hline \\[-1.8ex] 
 & \multicolumn{4}{c}{\textit{Dependent variable:}} \\ 
\cline{2-5} 
\\[-1.8ex] & \multicolumn{4}{c}{$\Delta\ln Y$} \\ 
 & \multicolumn{2}{c}{Annual} & \multicolumn{2}{c}{Quarterly} \\ 
\\[-1.8ex] & (1) & (2) & (3) & (4)\\ 
\hline \\[-1.8ex] 
 Constant & 0.017 & 0.028 & 0.007 & 0.009 \\ 
  & (0.009) & (0.010) & (0.008) & (0.005) \\ 
  $\Delta r$ & 1.935 & 1.682 & 2.424 & 1.414 \\ 
  & (0.367) & (0.303) & (1.029) & (0.475) \\ 
  $\Delta\ln M$ & 0.588 & 0.521 & 0.191 & 0.423 \\ 
  & (0.115) & (0.116) & (0.513) & (0.237) \\ 
 \hline \\[-1.8ex] 
Recession dummies & No & Yes & No & Yes \\ 
Quarterly dummies & No & No & Yes & Yes \\ 
Observations & 78 & 78 & 84 & 84 \\ 
R$^{2}$ & 0.483 & 0.645 & 0.207 & 0.518 \\ 
\hline 
\hline \\[-1.8ex] 
\end{tabular} 
\caption{Nominal Gross Output Growth, Interest-Rate Changes, and Credit Growth}
\label{tab:regress}
\begin{minipage}{\textwidth}\small
\textit{Notes:} Data are sourced from FRED and BEA. Columns (1)--(2) use the annual sample: 1948--2025; columns (3)--(4) use the quarterly sample. Parentheses report Newey-West standard errors with one annual lag or four quarterly lags. The period-average annual yield $r$ uses \textsc{ltgovbd} before March 1953 and \textsc{gs10} afterward. $M$ is year-end or quarter-end \textsc{tcmdo}. Annual $Y$ uses BEA gross output of all industries; quarterly $Y$ uses \textsc{goai} from FRED. Each recession dummy equals one when an observation includes a \textsc{usrec} recession month, from the month after the NBER peak through the trough. Quarterly specifications also include seasonal dummies to control seasonality.
\end{minipage}
\end{table}

\subsection{Summary}

The finance--value identity links interest rates, credit, prices, and gross production value. It explains why declining rates can accompany rising leverage, why growth can coexist with deflation, and why higher rates need not eliminate inflation. Each outcome depends on financing growth relative to value growth. The theory is supported by the empirical evidence from the United States.

It must be stressed, though, that the regressions only show that the theory is not at odds with the data; they do not validate it or rule out competing explanations. A theory's usefulness, however, does not depend on its goodness-of-fit with the data, but on its ability to explain history and advise policy choices. The next section addresses those tasks.

\section{Policy and Growth Regimes}

Now allow the value chain to expand over time by adding a higher-valued stage:
\[
V'=V+v_{N+1}.
\]
Higher value creation can emerge in different ways. It might involve a revolutionary product built on existing technologies, such as computers adopted by households and businesses. It might involve a service, such as YouTube subscriptions, with the potential to reach billions of users. A cost-reduction technology can create value by making a product once confined to laboratories and a wealthy few accessible to the wider population. Value creation need not necessarily involve technological breakthroughs or productivity improvement. Financial services become more valuable in a boom market, as more people participate in trading. 

Value creation is not a purely technological process. It is a social process, in which the market and policy environment shapes the value scale. As an example, higher education is the source of higher value creation in the US, but not equally so elsewhere. From 2000 to 2020, the public college tuition fee in the US increased by about 164\%; in the meantime, the general price level increased by about 50\%. During the same period, the public college tuition fee in China largely remained stable despite the general price level increasing by about 58\%.\footnote{The U.S. college tuition is from the \href{https://data.bls.gov/timeseries/CUUR0000SEEB01}{BLS series}. For China's college tuition, see a statement from the \href{https://www.moe.gov.cn/jyb_xwfb/s271/201408/t20140821_174052.html}{Ministry of Education}.} This is certainly not because US college tuition provides higher marginal utility than Chinese college tuition. It is the institutions, policy choices and public consensus that shape this result.

While value is a slowly evolving social construct, policy choices can change the general price level more rapidly, causing inflation or deflation. For better exposition, let $R=1+r$ denote the gross interest rate. Taking growth rates of \Cref{eq:rmpv-identity} gives
\begin{equation}
\hat R+\hat M=\hat p+\hat V.\label{eq:rmpv-growth}
\end{equation}
Thus, inflation occurs when growth in credit inclusive of interest exceeds value growth, while deflation occurs when it falls short. If value creation is regarded as a gradual process, then the policy mix that determines credit conditions and interest rates drives changes in the price level.

The remainder of this section examines the outcomes under different policy regimes. The joint movements of credit and interest rates, relative to value growth, help distinguish four regimes: deflationary growth, inflationary growth, reflationary growth, and deleveraging recessions. These regimes describe paths of prices, production, and financing consistent with the identity, although the identity alone does not establish which policy causes a particular outcome.

\subsection{Deflationary Growth}

This is the case where the monetary base cannot be freely expanded, typically under a gold standard. Broad money can grow through deposits and bank credit, though this expansion is constrained by the gold stock as money is ultimately redeemable in gold. According to \Cref{eq:rmpv-growth}, when $M$ and $R$ are relatively stable, $\hat p\approx-\hat V$. More generally, growth is deflationary when $\hat M+\hat R<\hat V$. 

In US history, 1879–1896 is a typical example of deflationary growth, following the return to the gold standard after the Civil War. Credit growth was roughly on par with final output growth (probably lower than gross output growth), while interest rates declined slightly. This combination produced a scenario of deflationary growth.

Under the gold standard, the stock of gold largely determines how far money and credit can expand and thus influences whether growth is inflationary or deflationary. By contrast, 1896–1914 was a period of increasing gold production, which allowed faster growth of money and credit and contributed to mild inflation.

Perhaps contrary to many economists' beliefs, deflation does not necessarily imply slow growth. As \textcite[93]{friedman1963} concluded, ``generally declining or generally rising prices had little impact on the rate of growth.'' Nevertheless, the process is painful for borrowers because the nominal amount they repay is worth more in real goods.

\subsection{Inflationary Growth}

This is typically the case under war finance, which combines state money creation, rapid debt growth, and competition for resources. Interest rates tend to rise, although wartime controls can restrain them. Inflation follows when
\[
\hat M+\hat R>\hat V.
\]
It can also occur in peacetime during a strong nominal boom, when private or public credit expands rapidly, possibly because of extraordinary business opportunities or strong government stimulus.

This relationship provides an interpretation of the two world wars, the 1970s, and the post-pandemic era. Large fiscal deficits and rising interest rates coexist with persistent inflation when credit growth remains strong relative to value growth.

Fighting inflation is difficult in this case. High interest rates restrain inflation through this identity only if they are associated with a sufficient slowdown in credit growth. Credit need not necessarily contract in level, but financing inclusive of interest must grow more slowly relative to value. This helps explain why, during the Volcker tightening of 1979–1982, interest rates had to be raised sufficiently high to rein in inflation.

\subsection{Recovery}

The period after the GFC and before the pandemic largely fits this regime: monetary authorities pursued low, stable inflation, commonly expressed as a 2\% target. Interest rates declined over the longer run to support full employment, despite cyclical fluctuations. Credit was largely free to expand or contract in response to demand.

In this stylized case, $\hat p\approx2\%$, value growth $\hat V$ is modest but persistent, and $\hat R<0$ over the long run. \Cref{eq:rmpv-growth} then implies
\[
\hat M>\hat p+\hat V.
\]
Financing therefore grows faster than nominal gross output. A rising debt-to-GDP ratio can approximate this pattern when nominal GDP tracks nominal gross output closely enough. The interpretation advanced here is that modest value growth is supported by increasing leverage, a pattern that can be described as ``debt-driven growth.'' Such a pattern often follows economic upheaval, as illustrated by the New Deal recovery, the expansion of the 1980s and 1990s following stagflation, and the post-GFC recovery before the pandemic.

\subsection{Recessions}

Recessions typically involve a deleveraging process that combines falling debt, $\hat M<0$, and falling interest rates, $\hat R<0$, with business bankruptcies and weak demand. This process produces deflation when $\hat M+\hat R<\hat V$. If gross output value is declining, this condition holds when credit inclusive of interest falls faster, typically when policy support is insufficient. The Great Depression of 1929--1933 is a typical example of a deflationary recession.

A recession need not be deflationary, however, if policy support is sufficient to ensure that $\hat M+\hat R \geq \hat V$. The policy response to the GFC illustrates how substantial monetary and fiscal support can limit deflationary pressures. This interpretation is consistent with the argument of \textcite[pp.~300--301]{friedman1963} that the monetary collapse during the Great Depression reflected the Federal Reserve's policy failures rather than the impotence of monetary policy.

\subsection{Summary}

\Cref{tab:history} summarizes historical episodes and changes in the relevant variables that are broadly consistent with the analysis above. The period divisions primarily reflect monetary regimes and wartime conditions. For example, whether the dollar is pegged to gold makes a substantial difference to the credit base and interest-rate determination. Wars also mark distinct regimes because governments often control prices and interest rates, such as the interest-rate controls spanning 1942--1951.

Some remarks are necessary when dealing with historical data. I use the \textcite{jorda2017} database for historical data. $M$ is approximated by private bank loans and public debt before 1946. From then on, it is approximated by All Sectors Debt Securities and Loans from the \textit{Financial Accounts of the United States}. Gross output across all industries is not available for the full historical sample, so I use real GDP as $Y$ throughout.

\begin{table}[!htbp]
\centering
\setlength{\tabcolsep}{4.5pt}\begin{tabular}{llrrrr}
\toprule
Period & Historical episode & $\Delta r$ & $\Delta\ln M$ & $\Delta\ln p$ & $\Delta\ln Y$ \\
\midrule
1880--1896 & Gold-standard deflation & -0.02 & 3.06 & -1.23 & 2.75 \\
1897--1913 & Gold-standard mild inflation & 0.05 & 5.97 & 0.96 & 4.29 \\
1914--1921 & WWI boom and bust & 0.02 & 14.01 & 7.40 & 1.43 \\
1922--1928 & Interwar expansion & -0.11 & 2.04 & -0.65 & 4.50 \\
1929--1933 & Great Depression & -0.09 & -5.03 & -5.48 & -5.53 \\
1934--1941 & Reflation and uneven recovery & -0.13 & 5.40 & 1.54 & 7.51 \\
1942--1951 & WWII and postwar rate control & 0.01 & 15.42 & 5.68 & 4.11 \\
1952--1971 & Bretton Woods expansion & 0.18 & 6.90 & 2.22 & 3.54 \\
1971--1981 & Gold-window closure and Great Inflation & 0.60 & 11.12 & 7.32 & 3.10 \\
1982--2007 & Disinflation and rising leverage & -0.36 & 8.91 & 3.01 & 3.03 \\
2008--2019 & Financial crisis and low-rate recovery & -0.21 & 2.69 & 1.77 & 1.54 \\
2020--2025 & Pandemic, inflation, and tightening & 0.36 & 6.18 & 3.84 & 2.35 \\
\bottomrule
\end{tabular}
\caption{Annualized Changes in Key Indicators across Historical Episodes}
\label{tab:history}
\begin{minipage}{\textwidth}\small
\textit{Notes:} Data are sourced from the \textcite{jorda2017} Macrohistory Database (JST) and FRED. $r$ is the long-term Treasury yield, $p$ is the consumer price index, and $Y$ is real GDP. $M$ is measured using the sum of private loans and public debt from JST before 1946 and FRED total debt of all sectors (\textsc{tcmdo}) thereafter. Data for the latest period are drawn from FRED. The table reports annual interest-rate changes in percentage points and annual log growth rates for the other variables, averaged over the years in each episode.
\end{minipage}
\end{table}

What would be the best policy mix? If deflationary growth is considered unacceptable, the target is $\hat p\geq0$. If excessive debt growth is also undesirable, an ideal target is $\hat M\leq\hat V$. \Cref{eq:rmpv-growth} then implies $\hat R\geq\hat p$: growth in the gross interest rate must at least match inflation. 

Yet jointly restraining inflation and debt growth can be highly restrictive. If $\hat p\approx0$ and $\hat M\approx0$, the identity requires $\hat R\approx\hat V$. The gross interest rate must rise at the rate of gross value growth; this would imply prohibitively high rates. 
Good policy requires a balance among these objectives, depending on the circumstances and policymakers' priorities. The theory makes its trade-offs explicit.

\section{Distribution Issues}

This section examines distribution as the value chain grows. Rearranging \Cref{eq:zero-profit} gives
\begin{align}
w_i&=\frac{pv_i}{1+r}-pv_{i-1}\nonumber\\
&=\frac{p(\Delta v_i-rv_{i-1})}{1+r}.\label{eq:wage}
\end{align}
For stages $i\geq 2$, viability requires $r\leq\Delta v_i/v_{i-1}$, including the zero-wage boundary. For stage 1, $w_1=pv_1/(1+r)$.  Thus, the providers of the original means of production are always rewarded. Wages of later stages depend on the value a stage creates net of the interest burden of financing intermediate inputs from the previous stage.

Total wages are
\begin{align}
W&=\frac{pv_1}{1+r}+\sum_{i=2}^N\left(\frac{pv_i}{1+r}-pv_{i-1}\right)\nonumber\\
&=\frac{pV}{1+r}-p(V-v_N)\nonumber\\
&=pv_N-\frac{rpV}{1+r}\nonumber\\
&=pv_N-rM.\label{eq:total-wage}
\end{align}
Total wages equal final-stage value minus interest income. More financing and higher interest charges direct more income toward interest earners and leave less for wages.

Given that economic progress is defined as greater value creation, living standards should not be measured by per capita purchasing power over quantities of goods, but by how much of the value created can be enjoyed by the population. If the economy produces more higher-valued exquisite products that are less affordable to ordinary wage earners, the majority of the population does not benefit from this economic progress. So define wage earners' purchasing power over final-stage value as
\begin{align}
\Omega&=\frac{W}{pv_N}\nonumber\\
&=1-\frac{r}{1+r}\frac{V}{v_N}\nonumber\\
&=1-\frac{rM}{pv_N}.\label{eq:omega}
\end{align}
If $pv_N$ represents final output measured by GDP, $\Omega$ is labor's share of final output. It captures the distribution between wage earners and interest earners. The wage share declines as interest payments rise relative to final output. This interpretation, however, comes with an important qualification: $v_N$ does not necessarily represent final goods. So the purchasing power interpretation is more appropriate. Nonetheless, I use the wage-share interpretation in the following discussion, as the distribution between wage earners and interest earners is an essential aspect of social inequality.

\Cref{eq:omega} also shows that the shape of the value chain matters, especially the ratio $V/v_N$. At a given interest rate, a constant ratio $V/v_N$ preserves the wage share, while a rising ratio reduces it.

The following discussion considers different shapes of the value chain. Readers shall be reminded again that value growth is not an exogenous process driven only by technological progress. Rather, it reflects a social process influenced by institutions, social norms, and policies. Consider the real estate market: an elastic fiat monetary system and mortgage policies can support housing prices far above what most wage earners can afford. Such price increases would not be possible under a fixed monetary base or under institutions that restrict the use of real estate as collateral for borrowing. The shape of the value chain therefore reflects social and policy choices.

\subsection{Linear Value Growth}

Let $v_i=ki$ for $i\geq1$, with $k>0$ and $N\geq2$. Then $\Delta v_i=k$ for $i\geq 2$ and $v_1=k$. Both cases give
\[
w_i^L=\frac{pk[1-r(i-1)]}{1+r}.
\]
The full chain is viable when $r\leq1/(N-1)$. The wage share is
\begin{align*}
\Omega^L&=1-\frac{r}{1+r}\frac{\sum_{i=1}^Nki}{kN}\\
&=1-\frac{r}{1+r}\frac{N+1}{2}.
\end{align*}
If $r=0$, the model resembles a socialist economy in which wages are perfectly equal and resources are allocated without capitalist interest income. Then $w_i^L=pk$ and $\Omega^L=1$. Wages are identical across stages, and their aggregate purchasing power covers all final output.

If $r>0$, wages decline linearly as the value chain grows, because constant value increments cannot overcome the accumulating financing burden. A long chain therefore requires a very low interest rate to remain viable. The wage share also falls as the chain lengthens, as interest absorbs more and more of the value created.

\subsection{Exponential Value Growth}

Consider a constant proportional value increment at stages $i\geq2$:
\[
\frac{v_i}{v_{i-1}}=1+g,\qquad i=2,\ldots,N,\quad g>0.
\]
Starting from $v_1>0$, the chain is
\[
v_i=v_1(1+g)^{i-1},\qquad i=1,\ldots,N.
\]
Gross value is
\[
V=\sum_{i=1}^N v_i=\frac{v_1[(1+g)^N-1]}{g}.
\]
The first stage has no purchased intermediate input. Its wage is $w_1^E=pv_1/(1+r)$. At later stages,
\begin{align*}
w_i^E&=\frac{p(\Delta v_i-rv_{i-1})}{1+r}\\
&=\frac{p(g-r)v_{i-1}}{1+r},\qquad i=2,\ldots,N.
\end{align*}
For nonnegative interest rates, viability requires $r\leq g$. Wages are positive when $r<g$ and zero when $r=g$, except for the suppliers of the original means of production. The aggregate wage share is
\begin{align*}
\Omega^E&=1-\frac{r}{1+r}\frac{v_1[(1+g)^N-1]/g}{v_1(1+g)^{N-1}}\\
&=1-\frac{r}{1+r}\frac{1+g-(1+g)^{1-N}}{g}\\
&\longrightarrow\frac{g-r}{g(1+r)}\qquad\text{as }N\to\infty.
\end{align*}
If $r=0$, then $w_1^E=pv_1$ and $w_i^E=pgv_{i-1}$ for $i\geq2$. With no interest charges, wage earners receive all value created as $\Omega^E=1$. However, wages increase exponentially as the value grows; wage inequality is worsening.

If $0<r<g$, wages also rise exponentially, but not as fast as before. The wage share falls toward a positive limit. Thus wage earners preserve their share against interest earners. This is because the growing upper-stage wage bill offsets part of the accumulation of interest claims. Unlike a linear chain, which reaches a finite viability limit at a fixed positive rate, this chain permits continued expansion. This coincides with the US postwar experience: ever-growing per capita GDP, rising wage inequality, and a relatively stable labor share.\footnote{See \textcite{piketty2014} for a discussion of rising inequality associated with the rise of ``supermanagers.'' Some studies report a declining labor share \parencite{autor2020}. However, once depreciation is subtracted from the calculation, the net labor share is relatively stable \parencite{bridgman2018}.}

The boundary $r=g$ has a different interpretation. All later-stage wages are zero. Only the providers of the original factors of production receive wage income; interest earners receive the rest of final output. Since $v_N=v_1(1+g)^{N-1}$,
\begin{equation}
\left.\Omega^E\right|_{r=g}
=\frac{w_1^E}{pv_N}
=\frac{pv_1/(1+g)}{pv_1(1+g)^{N-1}}
=(1+g)^{-N}.\label{eq:exponential-boundary}
\end{equation}
This share is positive for every finite chain. It approaches zero as $N$ increases because first-stage wages remain fixed while final-stage value grows exponentially.

For a long chain, the relative magnitudes of $r$ and $g$ determine the limiting wage share. A larger gap between $g$ and $r$ gives workers a larger share. This echoes \textcite{piketty2014}'s argument that $r>g$ allows capital to accumulate faster than labor income and contributes to inequality. The concepts differ, though: the present model's $g$ measures the rate of value growth, rather than final output growth. In this model, $r=g$ is viable with zero later-stage wages, while $r>g$ is not.

\subsection{Equalized Growth}

\textcite{piketty2014}'s argument that inequality is a choice raises a question: can growth be engineered to equalize wages? Value increments would need to offset accumulating financing costs without allowing upper-stage wages to dominate. By \Cref{eq:wage}, equal wages require
\[
\Delta v_i-rv_{i-1}=\bar v,
\]
for $i \geq 2$. Also let $v_1=\bar v$. $\bar v>0$ is a constant. Equal wages then satisfy
\[
w_i^Q=\frac{p\bar v}{1+r}.
\]
The implied value chain is
\[
v_i=\bar v\frac{(1+r)^i-1}{r}.
\]
The wage share becomes
\begin{align*}
\Omega^Q&=\frac{W}{pv_N}\\
&=\frac{Np\bar v/(1+r)}{p\bar v[(1+r)^N-1]/r}\\
&=\frac{Nr}{(1+r)[(1+r)^N-1]}.
\end{align*}
At $r=0$, the limiting chain is $v_i=i\bar v$, wages are $p\bar v$ at every stage, and $\Omega^Q=1$. For $r>0$, perfect wage equality does not protect labor's share: $\Omega^Q\to0$ as $N\to\infty$. This is because a longer value chain requires more financing. At any fixed positive interest rate, interest earners eventually claim almost all final-stage value under this assumption. Equal wages do not generate enough high wage earners to counter this accumulation. The trade-off is clear: exponential growth preserves labor's share while increasing wage inequality, while equalizing wages shifts economic gains away from wage earners toward interest earners.

\subsection{Prescription for Inclusive Growth}

My views on inequality and social welfare are in line with Piketty's. I do not believe that market allocation comes anywhere close to maximizing social welfare. It would be implausible to argue that redistributing a small portion of the wealth of the super-rich to those sleeping on the streets---an outcome that markets surely produce---would not improve social welfare. Inequality is a choice, and policies must actively ensure that growth is inclusive and benefits more than a small group of elites.

The theory nevertheless exposes a tension between capitalist growth and equality, especially between wage earners (workers) and interest earners (capitalists). Growth requires expanding financing, which inevitably directs a larger share of output toward interest earners. Equalizing wages across occupations and industries is not without a cost. Eliminating high wages will eventually shift the gains of economic progress to interest earners.

One possibility is growth without greater financing burdens: hold debt and interest rates stable while value expands. The theory implies deflationary growth in this case. It does not eliminate wage inequality, though. It increases workers' share if nominal final output rises relative to the fixed interest bill.

Inflationary growth does the opposite, as illustrated by the post-pandemic experience. Large-scale liquidity injections supported the economy, followed by high inflation and high interest rates. In this situation, workers are worse off in terms of both wage inequality and their share of output, despite economic prosperity. Nominal wage growth at the lower end of the distribution is offset by inflation; meanwhile, a higher \(rM\) reduces the wage share. Helicopter money intended to help people in financial distress can ultimately leave them worse off.

Another solution is to reduce production's dependence on financing, perhaps through a mixed capitalist-socialist system in which some production does not require finance. Taxation is another way of redistributing: tax interest earners and high-wage earners and transfer income to others.

\subsection{Summary}

Growth's distributional consequences depend on the shape of the value chain and the cost of financing it. Two forms of inequality must be distinguished: inequality among wage earners and inequality between wage earners and interest earners. Exponential value growth supports labor's share of final output while widening wage gaps. Engineered equal wages would allow interest income to absorb an increasing share of final output. Economic progress does not automatically mean a proportionate gain for workers.

\section{Conclusion}

This paper proposes a new framework for thinking about the macroeconomy from the perspective of value creation. Since Adam Smith, economics has placed productivity at the center of growth. The proposed framework departs from this tradition by viewing growth as movement up the value chain. It also departs from the mainstream methodology that models the economy as the equilibrium outcome of supply and demand curves derived from agents' optimal choices. I consider this approach unrealistic and unhelpful in many cases. Instead, the analysis takes a socially and institutionally given value scale as its starting point. The theory brings together value, production, finance, money, and distribution and helps explain different episodes in US economic history. It conveniently offers an explanation for several features of modern capitalism, including declining interest rates, rising debt-to-GDP ratios, and increasing inequality.

This framework has substantial implications. For developing countries seeking to catch up with the frontier, the theory suggests that focusing solely on improving productivity or technology is insufficient. Development also requires climbing the value chain: integrating into the global value chain, upgrading production, and producing higher-valued products. For the developed world, there is no limit to value growth as long as human creativity continues to generate new values. The framework also emphasizes the power of policy choices to shape economic outcomes. Policies can facilitate value creation and allow nominal growth to continue indefinitely. But there are trade-offs, especially concerning inflation and distributional equality.

The framework offers a new lens through which to examine growth. Consider, for example, the current enthusiasm surrounding the AI revolution. Much of the discussion focuses on how AI can improve productivity, particularly in cognitive work, and on concerns that it may replace workers with these skills. However, productivity improvements do not guarantee greater value creation. AI products themselves can extend the value chain. In other industries, such as film production, AI may reduce production costs without necessarily generating more or better films that society values. AI may change how cognitive work is performed, but human creativity and intelligence are not easily replaced in discovering what society considers valuable.

Finally, the work of this paper is largely conceptual rather than directly quantifiable. The value scale serves as an analytical tool, replacing utility optimization and supply and demand curves as the starting point of economic analysis. I hope this framework opens new perspectives, inspires new approaches to macroeconomics, and encourages further analysis along these lines.

\newpage
\section*{References}
\printbibliography[heading=none]

\end{document}